\documentclass[12pt]{iopart}

\usepackage{iopams}
\usepackage{graphicx}
\usepackage{url}
\expandafter\let\csname equation*\endcsname\relax
\expandafter\let\csname endequation*\endcsname\relax
\usepackage{amsmath}
\usepackage{setspace}

\renewcommand{\v}[1]{\ensuremath{\bm{#1}}} 
 
\newcommand{\pd}[2]{\frac{\partial #1}{\partial #2}} 
 
\let\baraccent=\= 
\renewcommand{\=}[1]{\stackrel{#1}{=}} 

\newcommand{\sech}{\, \mathrm{sech}}

\newcommand{\jac}{\mathcal{J}}

\newcommand{\vpar}{v_{\parallel}}
\newcommand{\vpars}{v_{\parallel,s}}

\usepackage{bm}

\begin{document}
\title[]{Predicting Negative Triangularity Edge Confinement with a Neutral-Coupled Full-$f$ Gyrokinetic Model}

\author{T N Bernard$^1$, A C D Hoffmann$^2$, M Francisquez$^2$, \\ F D Halpern$^1$, G W Hammett$^2$, A Marinoni$^3$, \\ C Chrystal$^1$, and G McKee$^4$}

\address{$^1$ General Atomics, San Diego, CA, USA}
\address{$^2$ Princeton Plasma Physics Laboratory, Princeton, NJ, USA}
\address{$^3$ University of California, San Diego, CA, USA}
\address{$^4$ University of Wisconsin, Madison, WI, USA}
\ead{bernardt@fusion.gat.com}

\vspace{10pt}

\begin{abstract}
Global full-$f$ gyrokinetic simulations coupled to a reduced neutral-fluid model reproduce key DIII-D experimental trends and identify mechanisms underlying improved confinement in negative triangularity (NT) configurations relative to positive triangularity (PT). Using Gkeyll, we simulate DIII-D inner-wall-limited discharges across the confined edge, scrape-off layer (SOL), and sheath boundaries, allowing the edge electric field to evolve self-consistently. Compared with an adaptive recycling-source model, the reduced neutral fluid model improves particle and energy balances and brings the simulated density and temperature profiles into closer agreement with measurements, with minimal additional computational cost. In NT, suppressed turbulent fluxes coincide with a broad, stationary zonal flow inside the last closed flux surface, whereas PT develops transient, fragmented shear layers. Spectral analysis indicates that modified fluctuation phase relationships and enhanced zonal shear both contribute to transport suppression despite broadly comparable fluctuation amplitudes. The larger co-current departure from a locally matched bi-Maxwellian in NT suggests a velocity-selective kinetic process that may help establish the edge radial electric field by modifying flows or current balance. These results identify kinetic and turbulent mechanisms associated with improved NT confinement within an experimentally established ELM-free regime and demonstrate the importance of coupling gyrokinetic plasma dynamics with neutral interactions across the confined plasma and SOL.
\end{abstract}

%
%
%
%
%

\section{Introduction}
Achieving net fusion energy gain requires robust tokamak regimes that maintain macroscopic edge stability while limiting extreme divertor heat loads. High-confinement regimes that are inherently free of edge-localized modes (ELMs) are therefore a major priority for reactor-scale magnetic confinement. Within such regimes, the multiscale boundary dynamics linking the last closed flux surface (LCFS), scrape-off layer (SOL), and plasma sheath govern both global confinement and impurity transport. The mechanisms that determine confinement and transport across these coupled regions, however, remain incompletely understood.

Standard high-confinement mode (H-mode) operation in conventional positive triangularity (PT) tokamaks relies on a steep edge pressure pedestal \cite{Wagner1982}. This sharp gradient is constrained by peeling-ballooning limits and can trigger periodic ELMs that produce large transient heat fluxes to plasma-facing components \cite{Leonard2014}. Negative triangularity (NT) configurations offer a compelling ELM-free operating scenario, demonstrating H-mode-like core confinement while avoiding ELMs over a broad range of conditions \cite{camenen2007, austin2019, Marinoni2019, Merlo2021, Nelson2022,happel2023overview,aucone2024experiments}. Experiments further show that NT plasmas can achieve better confinement and transport properties than comparable PT plasmas. Identifying the first-principles kinetic and fluid mechanisms responsible for these confinement differences is essential for assessing the extrapolation of NT operation to reactor-scale devices.

Predicting edge behavior requires self-consistently modeling the transition from closed to open field lines across the LCFS, SOL, and plasma-sheath interface \cite{stangeby1990plasma, Loizu2012}. This region introduces complex kinetic effects, such as sheath-driven flow regulation and particle trapping, that standard fluid approximations do not fully capture \cite{Batishchev1997, Tskhakaya2008}. Because zonal flows play a central role in turbulence regulation, a unified, first-principles model of the confined plasma, SOL, and sheath is needed to determine how boundary dynamics establish the radial electric field ($E_r$) and generate shear layers in NT configurations \cite{Diamond2005}.

The edge radial electric field has long been associated with finite-orbit-width effects and ion orbit loss. By preferentially removing ions from specific regions of velocity space, these processes can generate intrinsic rotation, ion temperature anisotropy, and a negative edge radial electric field that regulates turbulence and confinement \cite{Groebner1990,Shaing1992_IOL,Shaing1992_ErWidth,StoltzfusDueck2012,StoltzfusDueck2020,Zhu2023_IOL}. Recent work has demonstrated that differences in neoclassical orbit dynamics may contribute to the stronger edge electric fields and enhanced confinement observed in NT plasmas \cite{Nishimura2020,Wang2023,kramer2025}.

Furthermore, characterizing the interaction between neutrals and edge turbulence is important for predicting confinement in NT and other configurations. Although neutral dynamics strongly influence boundary transport, they are often omitted from core gyrokinetic simulations. Recycled neutrals undergo charge exchange and ionization, acting as particle sources and momentum and energy sinks. Through charge-exchange-mediated transport of toroidal angular momentum, neutrals can also modify intrinsic rotation and the edge radial electric field, with the response depending sensitively on neutral localization and plasma collisionality \cite{Fulop2001_NeutralEr,Fulop2002_NeutralPoloidal,Omotani2016_NeutralRotation}. These processes alter plasma profiles, drive intrinsic flows, and influence microinstabilities \cite{Wersal2015, coroado2022self,zholobenko2021, bernard2022, bernard2023, Wilkie2024_NF,eder2025}.

Edge transport has been studied extensively with drift-reduced fluid models \cite{Dudson2009,Giacomin2022,Stegmeir2018,bufferand2019}. These models can capture large-amplitude fluctuations and open-field-line losses, but the plasma boundary is not always sufficiently collisional for their fluid closures to remain valid. A global gyrokinetic treatment avoids this limitation while retaining finite-orbit-width effects and kinetic dynamics across closed and open field lines. Although $\delta f$ models successfully describe core turbulence, where fluctuations are small \cite{Kotschenreuther1995, Jenko2000, candy2018, Mandell2018}, edge simulations must evolve the background distribution because the fluctuations are large; this requirement motivates a full-$f$ formulation. Representative global full-$f$ gyrokinetic codes include XGC \cite{Chang2017}, JOREK-GK \cite{becoulet2026global}, GENE-X \cite{Frei2021}, and Gkeyll \cite{Hakim2020}.

Gyrokinetic studies have significantly advanced the understanding of transport in negative-triangularity plasmas. Global simulations with GENE and GENE-X have demonstrated reduced turbulent transport, the importance of nonlocal effects, and improved confinement in NT relative to PT \cite{Merlo2021,ulbl2025simulations,ulbl2026validation}. More recently, global full-$f$ gyrokinetic simulations with JOREK-GK in realistic X-point geometry have shown that NT develops stronger zonal flows through enhanced Reynolds stress, leading to reduced turbulent heat transport and improved confinement \cite{Becoulet2026}. Complementary XGC simulations have highlighted the roles of neoclassical orbit dynamics and finite-orbit-width effects in establishing edge flows and radial electric fields in NT plasmas \cite{ku2026difference}. These results motivate simulations that treat neoclassical orbit dynamics and turbulence simultaneously. Building on these efforts, the present work introduces a computationally inexpensive reduced neutral treatment within a global full-$f$ gyrokinetic framework to investigate improved confinement in NT plasmas.

In the present study, we use the global full-$f$ gyrokinetic solver in Gkeyll \cite{Hakim2020,francisquez2026,gkeyllWeb} to simulate PT and NT DIII-D inner-wall-limited (IWL) discharges. We employ the electrostatic, long-wavelength full-$f$ formulation \cite{francisquez2026} to model coupled closed- and open-field-line regions in limiter geometry, building on previous work \cite{Bernard2024,Hoffmann2026_AdptSrc,Hoffmann2026_HighThroughput}. The present calculations extend earlier Gkeyll simulations of these DIII-D discharges \cite{Bernard2024} by incorporating an adaptive source formulation \cite{Hoffmann2026_AdptSrc} and separate Miller equilibria fitted to the experimental PT and NT equilibria, including their different Shafranov shifts.

The central development of this work is a reduced neutral model that replaces the adaptive recycling source with ionization-driven particle fueling and includes ionization energy losses and charge-exchange momentum transfer. The neutral profile shape is prescribed, while its amplitude is determined from the global balance between recycling and ionization. This approach avoids the computational expense of six-dimensional kinetic neutrals \cite{bernard2022} and three-dimensional Monte Carlo coupling \cite{Shukla2025_LowRecycling}, while adding minimal computational overhead. The resulting simulations achieve improved particle and energy balances and closer agreement with measured DIII-D plasma profiles relative to previous simulations.

Using this framework, we first assess how the reduced neutral treatment changes plasma profiles, parallel losses, and global particle and energy balances relative to the adaptive recycling model. We then compare the PT and NT configurations and examine their zonal shear, turbulent transport, and velocity-selective kinetic signatures.

The remainder of this paper is organized as follows. Section \ref{sec:model} outlines the gyrokinetic and neutral equations, and Section \ref{sec:setup} describes the simulation setup. Section \ref{sec:neutrals} compares a baseline scenario to a neutral-coupled simulation. Section \ref{sec:tri} compares PT and NT neutral-coupled simulations and includes a detailed study of the turbulence properties. Section \ref{sec:conclusions} provides a summary and outlines future research directions.

\section{Model equations} \label{sec:model}
This section summarizes the gyrokinetic and reduced neutral-fluid models, with additional details contained in \ref{app:model-eqns}. We employ the global full-$f$ gyrokinetic solver in Gkeyll under a long-wavelength (guiding-center) approximation in the electrostatic limit for strongly magnetized plasmas \cite{francisquez2026}. We evolve the distribution function $f_s(\v{R}, \vpar, \mu)$ of plasma species $s$, where $\v{R}$ is the three-dimensional guiding-center coordinate vector, $\vpar$ is the velocity-space coordinate parallel to the magnetic field, and the magnetic moment $\mu = mv_\perp^2/(2B)$ defines the perpendicular velocity-space coordinate. The gyrokinetic equation governing this evolution is given in conservative form by
\begin{equation}
    \pd{\jac_s f_s}{t} + \nabla \cdot (\jac_s\dot{\v{R}}f_s) + \pd{}{\vpar}(\jac_s \dot{v}_\parallel f_s) = \jac_s C[f_s] + \jac_s S_s, \label{eq:gk-eqn}
\end{equation} where $\jac_s$ is the phase-space Jacobian of the transformation from particle coordinates to gyrocenter coordinates. Collision terms can include a Dougherty model of multi-species plasma collisions \cite{Francisquez2020lbo,francisquez2022improved} and plasma-neutral interactions, such as ionization, charge exchange, and recombination \cite{bernard2022}. For details of the collision operator, see \ref{app:coll}, and for neutral collision terms, see \ref{app:neutrals}. The source term $S_s$ includes adaptive sources that model fluxes from the core region and recycling processes \cite{Hoffmann2026_AdptSrc}, which are described in \ref{app:adpt-src}.

The gyrokinetic quasineutrality equation is given by
\begin{equation}
    -\nabla \cdot \sum_s \frac{m_s n_{s0}}{B_0} \nabla_\perp \phi = \sum_s q_s n_s = \sum_s q_s \int_{\Omega_{\v{v}}} d\v{v} \, \jac_s f_s, \label{eq:poisson}
\end{equation}
where $m_s$ is the species mass, $n_{s0}$ is a reference density, $B_0$ is a reference magnetic field magnitude, and $n_s$ is the species density. The left-hand side represents the linearized polarization term that retains lowest-order finite-Larmor-radius effects.

The simulations use a non-orthogonal, field-line-following coordinate system $(x, y, z)$ to represent the radial, binormal, and parallel directions, respectively. We apply standard twist-and-shift boundary conditions to the distribution function and electrostatic potential in the parallel direction on closed field lines \cite{Beer1995, francisquez2024}. On open field lines, conducting sheath boundary conditions reflect low-energy electrons, while ions stream freely out of the domain \cite{shi2017gyrokinetic}. Radial boundaries are absorbing, and the binormal direction is periodic.

We model the equilibrium magnetic flux surfaces using the following Miller parametrization
\begin{eqnarray}
R(r,\theta) &=& R_{axis} - \alpha_s \frac{r^2}{2R_{axis}} + r \cos(\theta + \sin^{-1}\delta\sin \theta), \label{eq:millerR} \\
Z(r, \theta) &=& Z_{axis} + r \kappa \sin \theta, \label{eq:millerZ}
\end{eqnarray}
where $\delta$ is the triangularity, $\kappa$ is the elongation, and $\alpha_s$ is the flux-surface compression factor related to the Shafranov shift. We calculate the minor radius $a$ self-consistently from Eq.~\ref{eq:millerR} using $R(a, 0) = R_{LCFS}$, which ensures that $R_{LCFS}$ in the simulation geometry accurately matches the experimental value. This Miller parametrization is valid for small inverse aspect ratios and neglects higher-order shaping parameters. While one can incorporate higher-order Miller models \cite{Turnbull1999,Riva2017,Arbon2020} by redefining the $R, Z$ mapping in the input file, such extensions lie outside the scope of the present work.

\subsection{Reduced fluid neutral model}
\label{sec:neut-model}
While previous simulations in simplified configurations incorporated fully kinetic neutrals \cite{bernard2022,bernard2023}, this work employs a reduced neutral-fluid model to lower computational overhead and accelerate simulations. The model is formulated in terms of the neutral mass density $\rho$, momentum density $\rho \v{u}$, and energy density $\mathcal{E}$. In the present implementation, however, the neutral drift velocity and temperature are held fixed, and only the density amplitude is updated dynamically. We scale the neutral density $n_n = \alpha n_n(t=0)$ according to the dynamically calculated coefficient
\begin{equation}
    \alpha = \frac{\mathcal{R} \int dA \, \Gamma_{i,wall}}{\int dV n_n(t=0) n_e \langle \sigma v_e \rangle_{iz}},
\end{equation}
where $\mathcal{R}$ is the user-defined recycling fraction (set to 0.98 for the present simulations), $\Gamma_{i,wall}$ is the ion particle flux incident on the walls or target plates, and $\langle \sigma v_e \rangle_{iz}$ is the ionization rate coefficient obtained from the OpenADAS atomic database \cite{OpenADAS}. This formulation serves as a pseudo-recycling boundary condition that enforces a global balance between the volume-integrated ionization rate (denominator) and the ion loss rate across the boundaries (numerator). The baseline spatial profile $n_n(\v{R},t=0)$ may be specified analytically or initialized from the output of a high-fidelity transport code.

We model ionization, charge exchange, and recombination interactions as collision terms, which are described in detail in \ref{app:neutrals}. The ionization and charge-exchange operators used here were previously verified against analytic results and benchmarked against DEGAS2 within the Gkeyll kinetic-neutral model \cite{bernard2022}. While a kinetic line-radiation model is available in Gkeyll \cite{Roeltgen2025}, we omit it from this work because radiation from main-ion neutrals is negligible under the plasma edge conditions of the considered discharges. Radiation from impurities may be more significant, but is not included in the present simulations. Future work will incorporate impurity species and radiation losses to assess their impact on edge transport.

\section{DIII-D simulation setup} \label{sec:setup}
This section describes the numerical setup for DIII-D inner-wall-limited (IWL) L-mode discharges with positive and negative triangularity, based on the experiments reported in \cite{Marinoni2019}. The magnetic equilibria and baseline kinetic profiles are derived from discharge \#171650 at 2800~ms for the PT case and discharge \#171646 at 2800~ms for the NT case; both employed electron cyclotron heating (ECH) as the primary auxiliary heating source. For the simulations, we assume that 1.5~MW of auxiliary heating power is deposited in the plasma in both configurations. We simplify the source model by partitioning this power equally between ions and electrons while neglecting impurity species and radiative losses. Over the simulated edge region (10~cm inside the LCFS), TRANSP calculations yield average effective charges of \(Z_{\mathrm{eff}}\simeq2.60\) for PT and \(Z_{\mathrm{eff}}\simeq3.10\) for NT. The corresponding net powers entering this region, inferred from the electron and ion energy-flux balances, are approximately 1.39~MW for PT and 1.78~MW for NT. Thus, the simulations are not expected to reproduce the experimental profiles exactly but instead provide a controlled comparison of the PT and NT configurations under similar source conditions.

\begin{figure}
    \centering
    \includegraphics[width=0.4\linewidth]{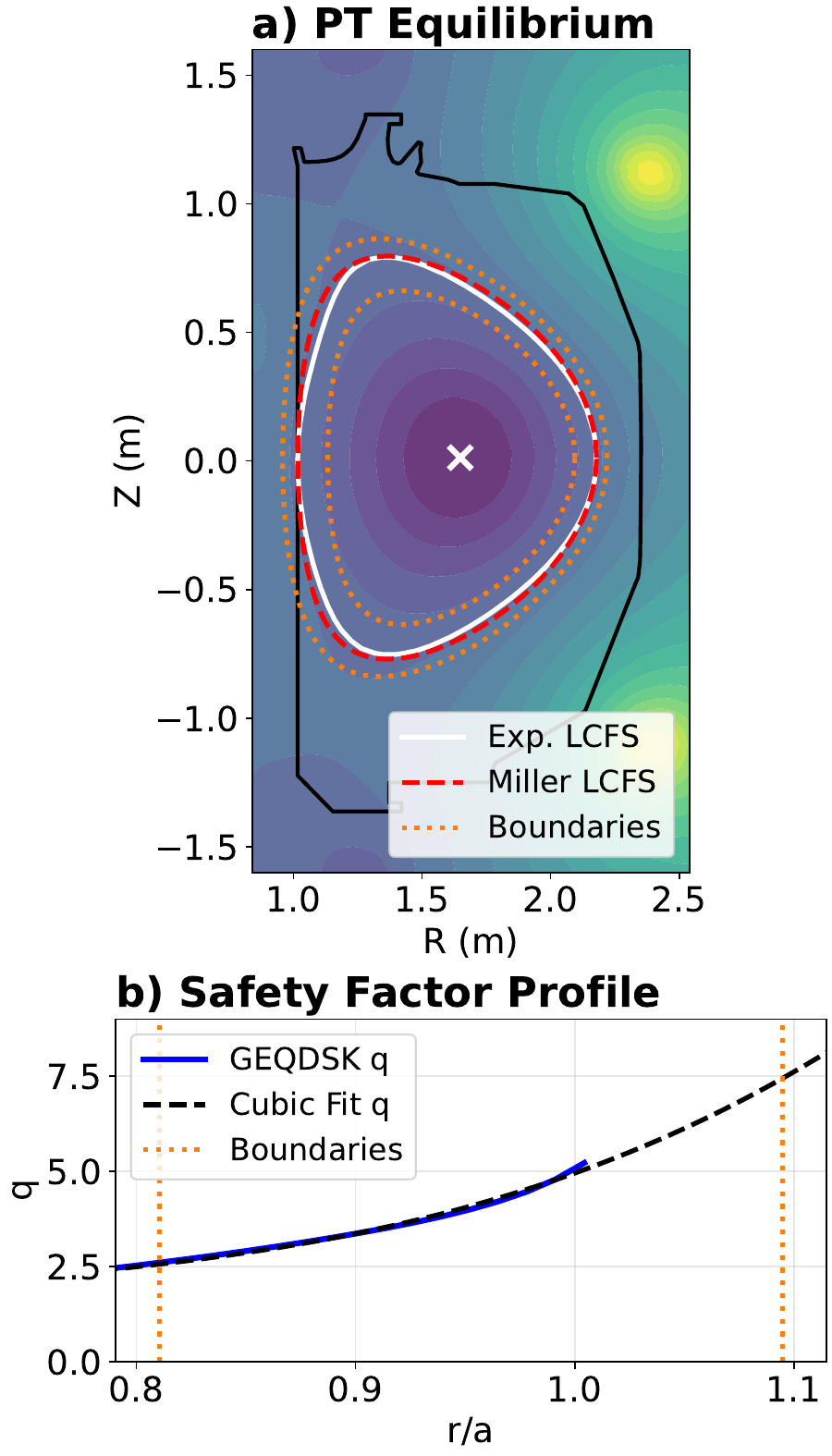}
    \includegraphics[width=0.4\linewidth]{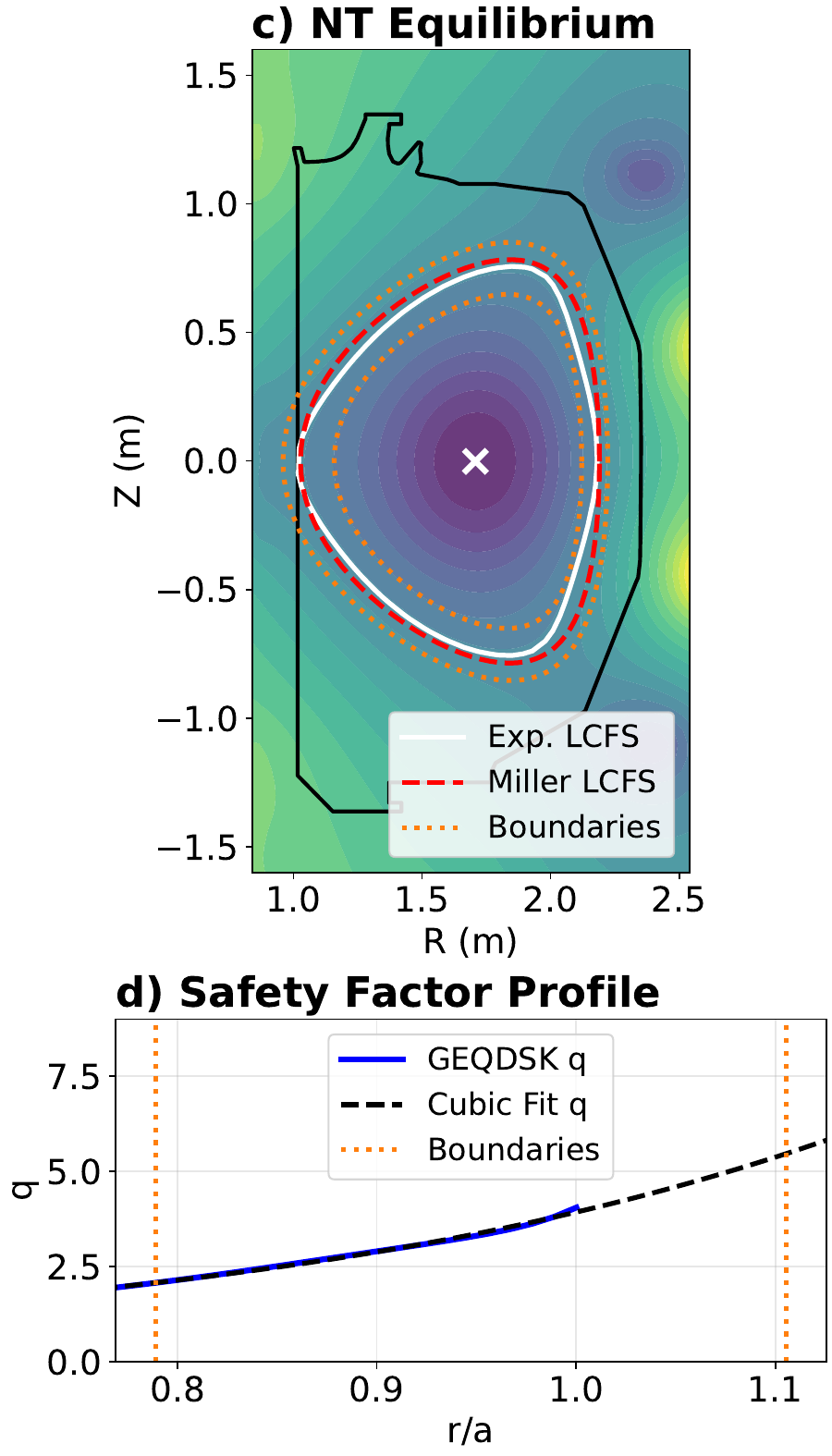}
   \caption{Poloidal magnetic flux contours (a, c) and edge safety factor profiles (b, d) for the PT (a)--(b) and NT (c)--(d) configurations. Contour maps display the experimental LCFS (solid white), analytical Miller LCFS (dashed red), and simulation boundaries (dotted orange). The radial $q$ profiles compare the raw GEQDSK data (solid blue) with the cubic fit (dashed black) over the normalized minor radius ($r/a$), with vertical dotted lines marking the simulation domain.}
    \label{fig:equil-q}
\end{figure}

The Miller parameters in Eqs.~\eqref{eq:millerR}--\eqref{eq:millerZ} are obtained by fitting to the LCFS from the experimental equilibrium reconstruction files. The elongation is $\kappa = 1.35$ and the triangularity magnitude is $|\delta| = 0.4$ for both configurations. The PT case has a Shafranov shift parameter of $\alpha_s = 0.5$, whereas the NT case has $\alpha_s = 1.0$. Because the experimental safety factor data only extend to the LCFS, a cubic fit was applied to the experimental $q$-profile in the edge region of interest ($r/a \geq 0.85$) and extrapolated into the SOL. Both the resulting Miller equilibria and the safety factor profiles are compared with the experimental data in Fig.~\ref{fig:equil-q}. The safety factor is lower in the NT configuration [Fig.~\ref{fig:equil-q}(d)], a property that generally holds when comparing NT to PT equilibria at fixed plasma current~\cite{zheng2024prospects}.

The simulations evolve a deuterium plasma with kinetic ions and electrons using realistic mass values. We specify a reference density $n_0 = 2 \times 10^{19}$~m$^{-3}$ and reference temperatures $T_{e0} = T_{i0} = 300$~eV. These reference parameters determine the Coulomb logarithm used to calculate the collision frequencies in the multi-species Dougherty collision operator. Because collisions often limit the explicit time step, we reduce the collision frequencies by a factor of two to accelerate the simulations. This reduction has a negligible effect on profile formation and turbulent transport. Full details of the collision operator implementation are provided in Ref.~\cite{francisquez2022improved}.

The reference magnetic field $B_0$ corresponds to the field strength at the center of the radial domain on the outboard midplane (OMP), yielding approximately 1.54~T for PT and 1.59~T for NT. This is used along with the reference temperatures to define the velocity-space grid extents. Additional relevant baseline parameters include the sound speed $c_s = \sqrt{T_e/m_i} \simeq 1.20 \times 10^5$~m/s and the ion sound gyroradii, $\rho_{s,PT} \simeq 1.64$~mm and $\rho_{s,\rm NT} \simeq 1.58$~mm.

The radial grid extends 10~cm inward from $R_{\rm LCFS}$ into the closed flux region and 5~cm outward into the SOL. The resulting 15~cm domain spans approximately 92~$\rho_s$ in PT and 95~$\rho_s$ in NT. The binormal extent is $L_y = 150\rho_s$, with $y \in [-L_y/2, L_y/2]$. The parallel direction is parameterized by the poloidal angle over a full poloidal turn, such that $z \in [-\pi, \pi]$. In parallel velocity space, the grid extends across $\vpars \in [-5v_{t0,s}, 5v_{t0,s}]$, where $v_{t0,s} = \sqrt{T_{s0}/m_s}$. To resolve a broader range of kinetic energies without increasing numerical resolution, we use a non-uniform velocity-space grid \cite{francisquez2026}. Specifically, a linear mapping is applied for $|\vpars/v_{t0,s}| \leq 2.5$ and a quadratic mapping for $2.5 < |\vpars/v_{t0,s}| \leq 5$. The magnetic moment grid spans $\mu \in [0, m_s(4v_{t0,s})^2/(2B_0)]$, also deploying a non-uniform quadratic coordinate mapping.

Gkeyll employs a discontinuous Galerkin method \cite{Juno2018,Hakim2020}. We use a linear polynomial basis ($p=1$) in configuration space, yielding 8 degrees of freedom per cell. For phase-space quantities, we implement a hybrid basis where the $\v{R}$ and $\mu$ variables use a linear basis, while $\vpar$ uses a quadratic basis \cite{francisquez2026}. This formulation provides 48 degrees of freedom per phase-space cell.

\subsection{Plasma initial conditions and sources} \label{subsec:setup-ic-src}
We initialize the plasma species with density and temperature profiles that approximate experimental measurements, accelerating relaxation toward a steady state. Despite differences between the experimental PT and NT profiles, we define identical initial conditions for both configurations to ensure a controlled comparison.

All simulations include an adaptive source at the inner radial boundary to model energy transport from the unsimulated core region. Because the discharges of interest use ECH exclusively, this source injects heat without introducing a net particle source. Experimentally, ECH power deposits energy preferentially into the electron channel. However, to establish a numerical baseline consistent with previous work \cite{Bernard2024,Hoffmann2026_AdptSrc}, we partition the input power equally between ions and electrons. Future studies will investigate the impact of more realistic asymmetric heating profiles.

We scale the input power to the flux tube domain via $P_{\rm in} = \mathcal{V}_f P_{\rm exp}$, where the volume fraction $\mathcal{V}_f = q_0 L_y/(2 \pi r_0)$ is approximately 0.303 for PT and 0.264 for NT. The direction of the toroidal magnetic field drives a downward ion $\nabla B$ drift, causing ions to exit the domain at $x_{\min}$ within the upper half of the poloidal plane. To model the compensating influx of ions drifting across the lower $x$-boundary, we position the ion core sub-source in the lower half of the $z$-plane. Conversely, electrons undergo $\nabla B$ drift in the opposite direction, so the electron core sub-source is placed in the upper $z$-plane.

To isolate the effect of the reduced neutral-fluid model, we establish a baseline simulation without explicit neutral interactions. Following Ref.~\cite{Hoffmann2026_AdptSrc}, this case includes an adaptive recycling source for ions and electrons in the region where ionization is expected. The source assumes that ions striking the limiter plates recycle and re-ionize instantaneously, thereby providing equal particle source rates for ions and electrons. We set the recycling source temperature to 10~eV. This added energy does not represent the ionization and charge-exchange energy losses included in the explicit neutral treatment. We determine the spatial distribution of the adaptive recycling source following Ref.~\cite{coroado2022self}.

We summarize relevant simulation parameters in table~\ref{tab:setup}. The transit-normalized collisionalities, $\nu_{ei}\tau_{\mathrm{tr},e}$ and $\nu_{ii}\tau_{\mathrm{tr},i}$, are evaluated at $r/a=1.01$ using $\tau_{\mathrm{tr},s}=L_c/v_{ts}$ and are averaged over the final $0.1~\mathrm{ms}$ of each simulation. These values characterize the parallel collisional regimes of the three cases: the electron collisionality ranges from below unity to greater than two, while the ion collisionality remains below unity in all cases. The complete geometry, initial conditions, and source definitions are available in the Gkeyll input files at \url{https://github.com/ammarhakim/gkyl-paper-inp/tree/master/2026_PPCF_NT_neut}.

\subsection{Neutral initial conditions} \label{subsec:setup-neut}
We prescribe an analytic neutral-density profile that approximates the steady-state balance between recycling and ionization. The objective is not to predict the neutral distribution from first principles but to provide an efficient reduced closure for the dominant neutral-plasma interactions in large-scale full-$f$ gyrokinetic simulations.

To construct this profile, we solve a steady-state diffusion equation and then fit an analytic function to the resulting solution. We solve for the neutral density $n_n(t=0)$ from
\begin{gather}
    \nabla \cdot (D_n \nabla n_n(t=0) ) - n_n(t=0) n_e \langle \sigma v_e \rangle_{iz} = 0, \\
    D_n = \frac{v_{ti}^2}{n_e \langle \sigma v \rangle_{cx}}, \label{eq:neut-diff}
\end{gather}
where $\langle \sigma v_e \rangle_{iz}$ and $\langle \sigma v \rangle_{cx}$ are the ionization and charge exchange reaction rates, respectively. In the curvilinear coordinate system, we assume the neutral density is axisymmetric, i.e., uniform in the binormal ($y$) direction. We apply Neumann boundary conditions at the inner radial boundary ($x_{\min}$), Dirichlet boundary conditions at the outer radial boundary ($x_{\max}$), periodic parallel boundary conditions in $z$ for the closed-field region ($x \leq x_{\rm LCFS}$), and parallel Dirichlet boundary conditions in the SOL ($x > x_{\rm LCFS}$). The outer radial Dirichlet boundary condition is not uniquely determined from first principles and therefore serves as a modeling parameter in constructing the initial neutral profile. The target plate boundary conditions along $z$ are written as
\begin{equation}
n_n v_{tn} = \mathcal{R} \Gamma_i,
\end{equation}
where $\Gamma_i$ is the outgoing ion particle flux and $\mathcal{R}$ is the recycling coefficient, set to 0.98. Here we assume $\Gamma_i = n_i c_s$. This system yields a matrix equation that we solve in a pre-processing step using the geometric data and plasma moments calculated from Gkeyll initial conditions.

\begin{figure}
    \centering
    \includegraphics[width=0.48\linewidth]{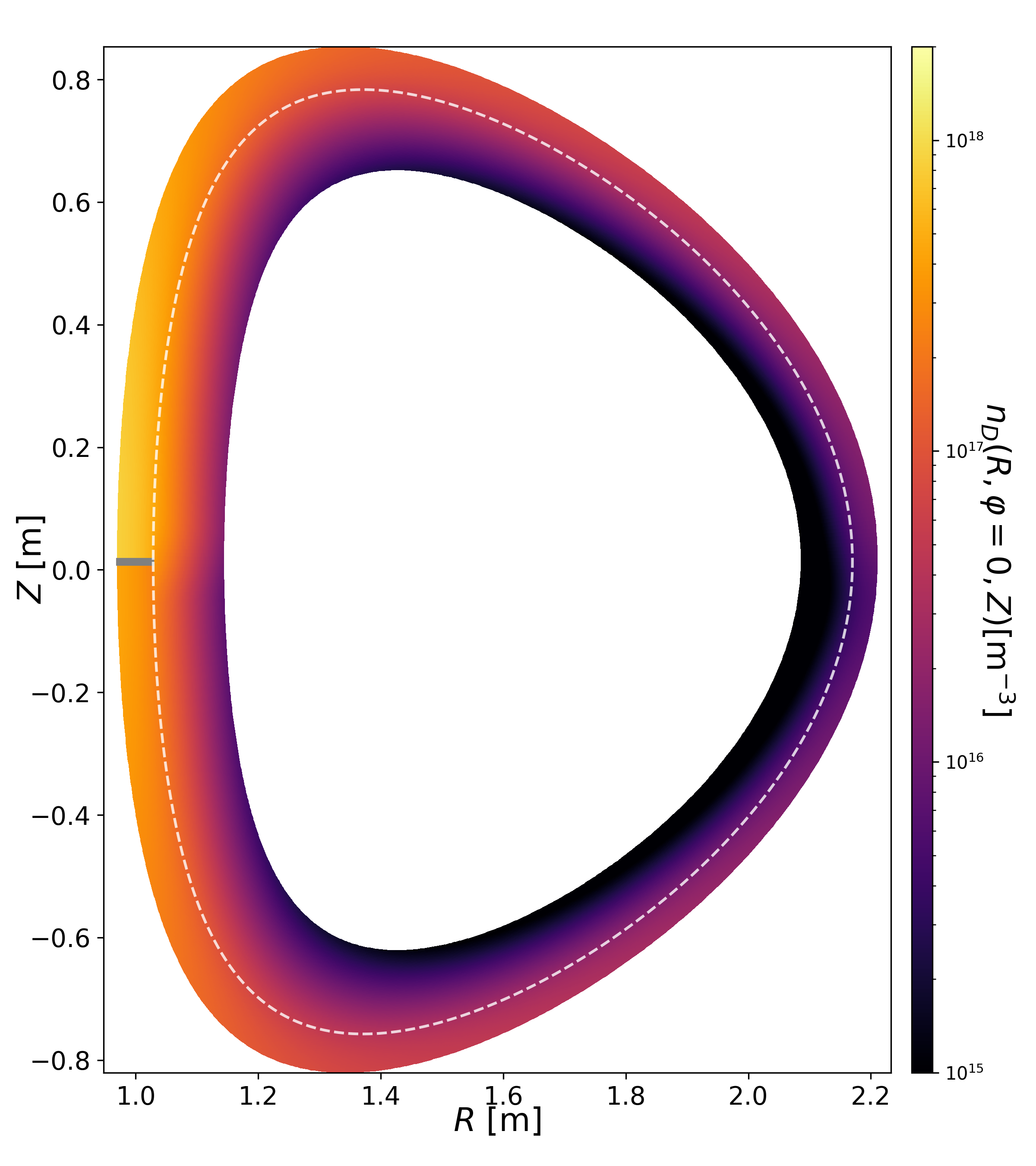}
    \includegraphics[width=0.48\linewidth]{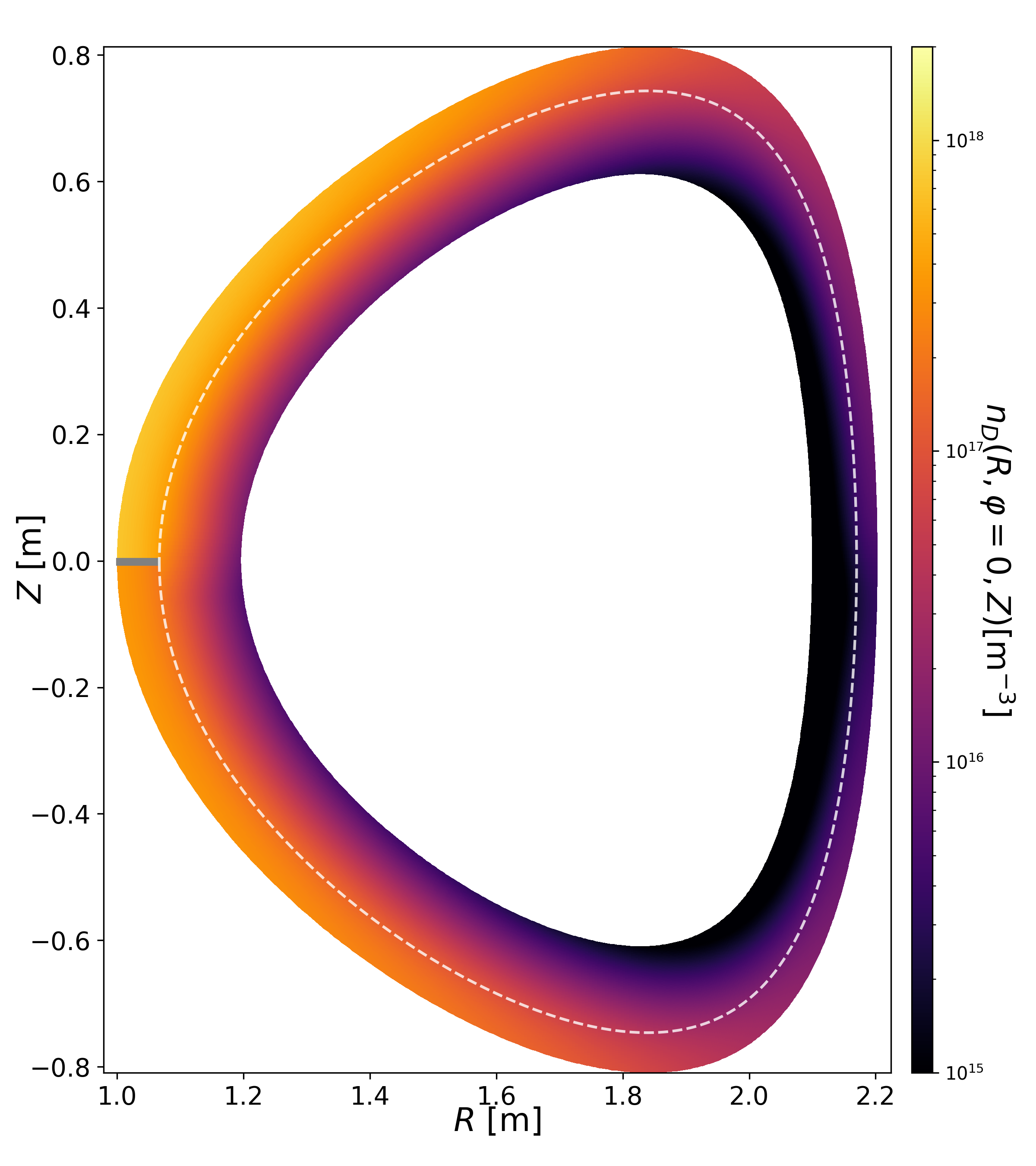}
    \caption{Neutral atomic deuterium density distributions after 2.5 ms for PT (left) and NT (right) configurations, with the LCFS indicated by the dashed white line.}
    \label{fig:neut-profile}
\end{figure}

The solution to equation \ref{eq:neut-diff} is used to initialize fitting parameters in the following analytic form for the initial neutral density:
\begin{equation}
n_n(x, z, t=0) = \begin{cases}
    n_{n0, \rm low} \left[ 1 - \tanh \left( \frac{x_{\text{LCFS}} - x}{\ell_{\text{rad}}} \right) \right] \left[ \sech^2 \left( \frac{z + L_z/2}{\ell_{\text{pol}}} \right) + 10^{-6} \right] & \text{for } z \le 0 \\
    n_{n0, \rm up} \left[ 1 - \tanh \left( \frac{x_{\text{LCFS}} - x}{\ell_{\text{rad}}} \right) \right] \left[ \sech^2 \left( \frac{z - L_z/2}{\ell_{\text{pol}}} \right) + 10^{-6} \right] & \text{for } z > 0.
\end{cases}
\end{equation}
The parameters $n_{n0,\rm up} = 2 n_{n0,\rm low}$, $\ell_{\rm rad} = 0.05$, and $\ell_{\rm pol} = 1.0$ were calibrated manually using low-resolution simulations to obtain near-steady-state particle balance while retaining the characteristic structure of the diffusion solution. A symmetric neutral distribution produced excessive differences between the upper and lower plasma profiles. We therefore imposed a poloidally asymmetric distribution qualitatively consistent with kinetic simulations of atomic and molecular neutral transport \cite{coroado2022self}. The resulting atomic-deuterium density distributions are shown in Fig.~\ref{fig:neut-profile}. Their magnitudes are consistent with available DIII-D measurements and edge-neutral modeling: atomic neutral densities near the outer-midplane separatrix have been inferred to be approximately $10^{15}$--$10^{16}\ {\rm m}^{-3}$ \cite{colchin2003,Sciortino2022}. Because no shot-specific neutral density measurements or Monte Carlo neutral calculations are available for these discharges, these comparisons establish order-of-magnitude consistency rather than quantitative validation.

Future work will replace the prescribed profile with a dynamically evolved neutral-fluid model or external coupling to kinetic neutral transport codes such as EIRENE, as in Ref.~\cite{Shukla2025_LowRecycling}. This will enable a more detailed study of the effect of shaping on neutral transport, which is outside the scope of the present work. Although these alternatives would describe neutral transport more self-consistently and require fewer free parameters, the present reduced neutral-fluid model captures the dominant recycling, ionization, and charge-exchange processes at negligible additional computational cost.

\begin{table}[h]
    \centering
    \begin{tabular}{|c|c|c|c|}
        \hline
        & PT base & PT neut & NT neut \\
        \hline
        Magnetic axis, $(R_{\text{axis}},Z_{\text{axis}})$ [m] & $(1.65,0.01)$ & $(1.65,0.01)$ & $(1.71,0.0)$ \\
        LCFS at OMP, $(R_{\text{LCFS}})$ [m] & 2.17 & 2.17 & 2.17 \\
        Shafranov shift, $\alpha_s$ & 0.5 & 0.5 & 1.0 \\
        Elongation, $\kappa$ & 1.35 & 1.35 & 1.35 \\
        Triangularity, $\delta$ & 0.4 & 0.4 & -0.4 \\
        Safety factor at separatrix, $q_{sep}$ & 5.1 & 5.1 & 4.0 \\
        Core power source  & yes & yes & yes \\
        Recycling particle source & yes & no & no \\
        Neutral interactions & no & yes & yes \\
        electron collisionality, $\nu_{ei}\tau_{\mathrm{tr},e}(r/a=1.01)$
        & $0.864$
        & $1.716$
        & $2.762$ \\
        ion collisionality, $\nu_{ii}\tau_{\mathrm{tr},i}(r/a=1.01)$
        & $0.156$
        & $0.406$
        & $0.345$  \\
        \hline
    \end{tabular}
    \caption{A summary of features of the simulations compared in this study. For normalized collisionalities, transit times are calculated as the field line connection length over the thermal speed, $\tau_{\mathrm{tr},s} = L_c/v_{ts}$.}
    \label{tab:setup}
\end{table}

\subsection{Resolution and computational costs} \label{subsec:setup-res-cost}

We first evolve the baseline PT case and the neutral-coupled PT and NT cases to 2~ms at low resolution, $(N_x, N_y, N_z) = (48, 36, 16)$. We then restart the neutral-coupled cases at high resolution, $(N_x, N_y, N_z) = (60, 60, 16)$, and evolve them for an additional 0.5~ms. All simulations share a velocity-space resolution of $(N_{\vpar}, N_\mu) = (12, 8)$.

We carried out the simulations on NVIDIA A100 GPU nodes on the NERSC Perlmutter supercomputer, using four nodes per simulation. The low-resolution PT simulations required 496 node-hours without neutrals and 630 node-hours with neutrals to reach 2~ms. The NT geometry imposed a more restrictive time-step limit, increasing the cost to approximately 868 node-hours. We therefore restricted the no-neutrals comparison to PT, and all NT simulations include the reduced neutral-fluid model. To complete an additional 0.5~ms, the high-resolution restarts required approximately 236, 310, and 435 node-hours for the PT baseline, neutral-coupled PT, and NT cases, respectively.

\section{Effects of neutral interactions} \label{sec:neutrals}
To study the effect of the reduced neutral-fluid formulation, we compare a baseline case with an adaptive recycling source but no neutral collisions (denoted ``PT base'' in figures) with a neutral-coupled case (denoted ``PT neut''). In the latter, ionization of the neutral species replaces the adaptive recycling source for the plasma species. Both high-resolution simulations were restarted from low-resolution states and evolved for an additional 0.5~ms, giving a total duration of 2.5~ms. Over the full simulation, the baseline particle inventory increases by 2.5\%, while its total energy increases by approximately 12\%, indicating excessive power input from the combined core and recycling sources. With neutrals, the particle inventory and total energy decrease by 3.8\% and 5.8\%, respectively, indicating improved global balance. Unless otherwise noted, all radial profiles are averaged over the final 0.1~ms and over the binormal direction $(\langle A \rangle_{y,t})$; for simplicity, these subscripts are omitted from plot labels.
\begin{figure}
    \centering
    \includegraphics[width=0.99\linewidth]{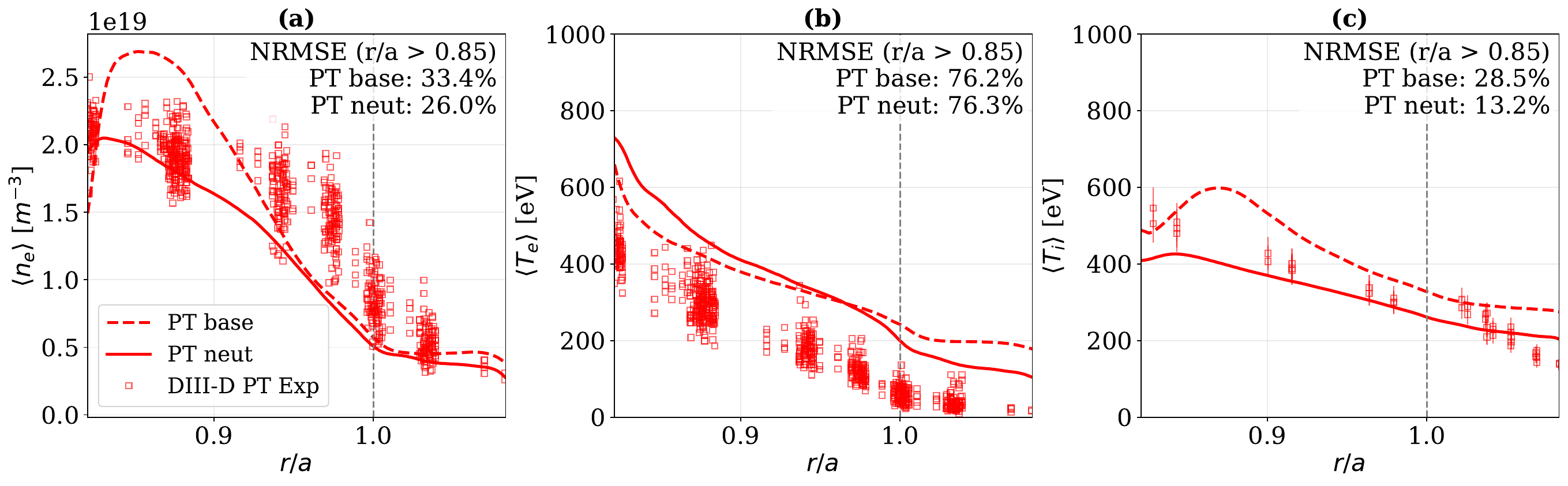}
    \caption{Time- and y-averaged radial profiles at the OMP of (a) electron density, (b) electron temperature, and (c) ion temperature for the PT configuration. Simulated profiles from the baseline case (dashed) and the case with neutral interactions (solid) are compared against DIII-D experimental measurements (scatter points). The LCFS is indicated by the vertical dashed line.}
    \label{fig:pt-neut-n-T-d3d}
\end{figure}

First, we compare the quasi-steady-state profiles with experimental measurements. Figures~\ref{fig:pt-neut-n-T-d3d}(a)--(c) display the plasma density, electron temperature, and ion temperature, respectively. Density and electron-temperature profiles are measured by Thomson scattering, while ion-temperature data are obtained from charge-exchange recombination spectroscopy, with C$^{6+}$ used as a proxy for the main-ion deuterium temperature. We extract the simulation values at the outboard midplane ($z = 0$) and average them over the binormal $y$-coordinate and the final 0.1~ms. The normalized root-mean-square error (NRMSE) between the simulated and experimental profiles is reported in each panel legend. It is calculated as $\lVert y_{\rm sim}(x_i)-y_{{\rm exp},i}\rVert_2/\lVert y_{\rm exp}\rVert_2$, where $y_{\rm sim}(x_i)$ is the linearly interpolated simulation value at the experimental measurement location $x_i$.

The neutral-coupled case replaces the adaptive recycling source with particle fueling through ionization and modifies parallel transport and boundary losses through charge exchange. These changes alter the global particle balance and allow the density profile to respond throughout the confined-edge domain, even though the neutral population remains concentrated near the plasma boundary. The baseline simulation overpredicts the inner-edge density, with an NRMSE of 33.4\%, compared with 26.0\% in the neutral-coupled case. The $T_e$ NRMSE is nearly identical in both cases. Both simulations overpredict $T_e$, likely because of simplifications in the source model and differences between the simulated and experimental power-deposition profiles. Nonetheless, agreement improves near the LCFS and in the SOL because ionization energy losses are modeled explicitly. The neutral-coupled case agrees well with the $T_i$ measurements from approximately $r/a=0.9$ to 1.05 and reduces the NRMSE by a factor of two relative to the baseline.

The neutral-coupled case matches all three profile gradients more closely than the baseline case, which is important for accurately modeling turbulent dynamics. Consistent with these profile changes, the density and potential fluctuations are lower when neutrals are included. As shown in Sec.~\ref{sec:tri}, the normalized fluctuation levels tend to exceed the measured values, so this reduction improves agreement with experiment.

\begin{figure}
    \centering
    \includegraphics[width=0.66\linewidth]
    {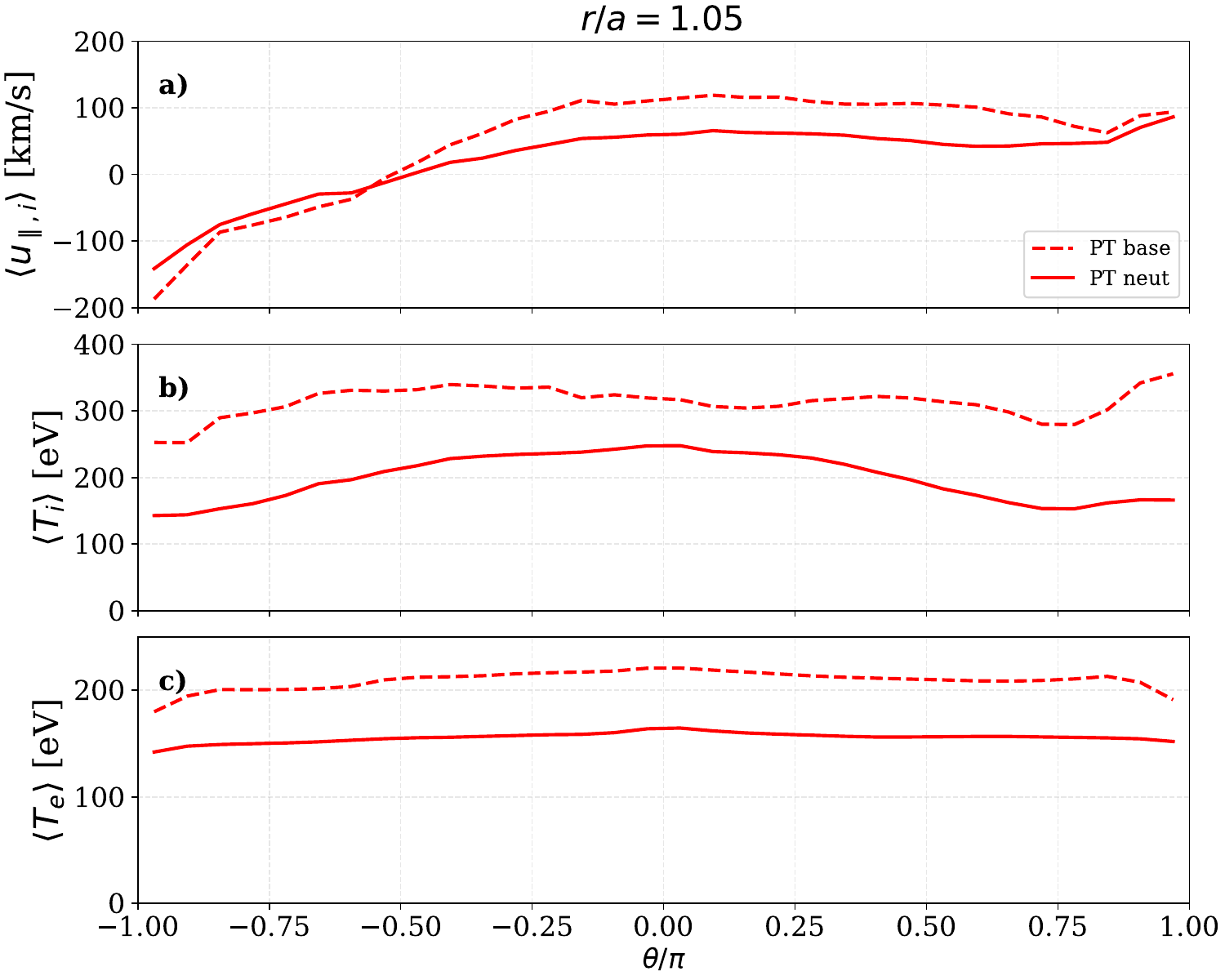}
    \caption{Parallel variations of (a) ion parallel velocity, (b) ion temperature, and (c) electron temperature in the SOL ($r/a = 1.05$) for the PT configuration. Profiles are plotted against the normalized poloidal angle $\theta/\pi$, comparing cases with (solid) and without (dashed) neutral interactions.}
    \label{fig:pt-neut-z-upar-Ti}
\end{figure}

Figure~\ref{fig:pt-neut-z-upar-Ti} displays the parallel variations of the ion parallel drift velocity, ion temperature, and electron temperature in the SOL at $r/a = 1.05$. Charge-exchange friction reduces the parallel-flow magnitude, while charge-exchange energy losses lower the ion temperature. The weaker parallel flows remove heat less effectively, producing a more strongly peaked parallel ion-temperature profile. Ionization losses also reduce the electron temperature. These trends align with previous simulations that used a fully kinetic neutral model \cite{bernard2023}.

Having established that the reduced neutral model yields a more physically consistent and experimentally relevant edge state, we use this formulation for the subsequent comparison of PT and NT configurations.

\section{Transport dynamics in positive and negative triangularity} \label{sec:tri}
This section compares the neutral-coupled PT and NT simulations. Both employ the same reduced neutral-fluid model and reaction physics, with profile parameters calibrated independently to establish steady-state particle balance in each magnetic configuration. We restarted these runs from the corresponding low-resolution simulations at 2~ms and evolved them on high-resolution grids for an additional 0.5~ms. During this interval, the NT particle inventory remained nearly constant while the total energy decreased by less than 2\%, confirming that the simulation maintains a statistically stationary state. (PT values are listed in Sec.~\ref{sec:neutrals}.) Unless otherwise noted, all profiles and turbulence statistics in this section are averaged over the final 0.1~ms of the high-resolution runs. Radial profiles are extracted at the OMP ($z=0$) and averaged over the binormal $y$-coordinate. We first compare the simulations with experimental measurements and then examine how negative triangularity modifies the turbulence, zonal flows, and kinetic edge dynamics.

\subsection{Comparison with experimental measurements} \label{subsec:tri-exp}
\begin{figure}
    \centering
    \includegraphics[width=0.99\linewidth]{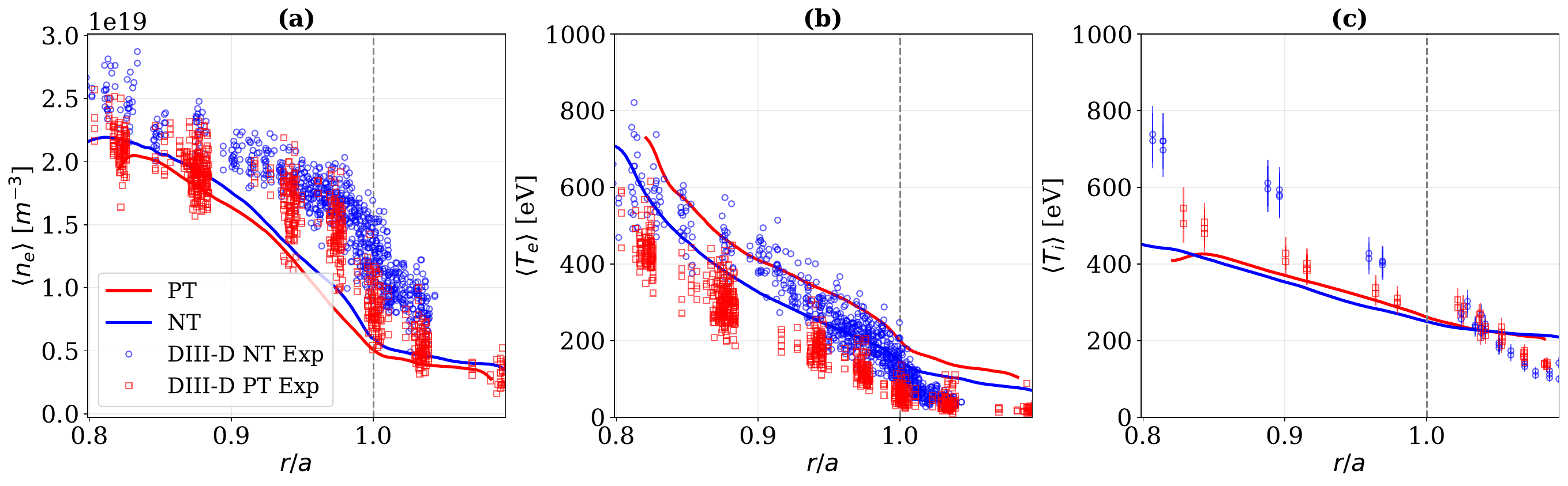}
    \caption{Time- and y-averaged radial profiles at the OMP of (a) electron density, (b) electron temperature, and (c) ion temperature for the PT (red) and NT (blue) configurations. Simulation profiles are plotted against normalized minor radius ($r/a$) and compared with DIII-D experimental measurements (scatter points), with the LCFS denoted by the vertical dashed line.}
    \label{fig:nt-pt-n-T-exp}
\end{figure}

Figure~\ref{fig:nt-pt-n-T-exp} compares the simulations with DIII-D measurements. The simulations neglect details of the experimental heating profiles, which may contribute to discrepancies with the measurements. Although they slightly underpredict the plasma density in panel (a), they reproduce the experimental trend of marginally higher density in NT than in PT. They do not, however, reproduce the measured temperature trends. The simulated NT electron-temperature profile agrees reasonably well with the measurements over much of the domain, whereas the PT simulation overpredicts $T_e$. Neither simulation reproduces the sharp experimental edge-$T_e$ structure. The simulated ion-temperature profiles are nearly identical, while the measurements show a higher edge $T_i$ in NT. TRANSP analysis indicates that the power deposited in the PT plasma was approximately \(20\%\) lower than in the NT plasma (see Sec.~\ref{sec:setup}), which may explain the disagreement in the temperature trends. A more realistic heating model that accounts for the experimental power-deposition profiles may improve agreement, but implementing such a model in flux-driven gyrokinetic simulations is beyond the scope of this work.

\begin{figure}
    \centering
    \includegraphics[width=0.8\linewidth]{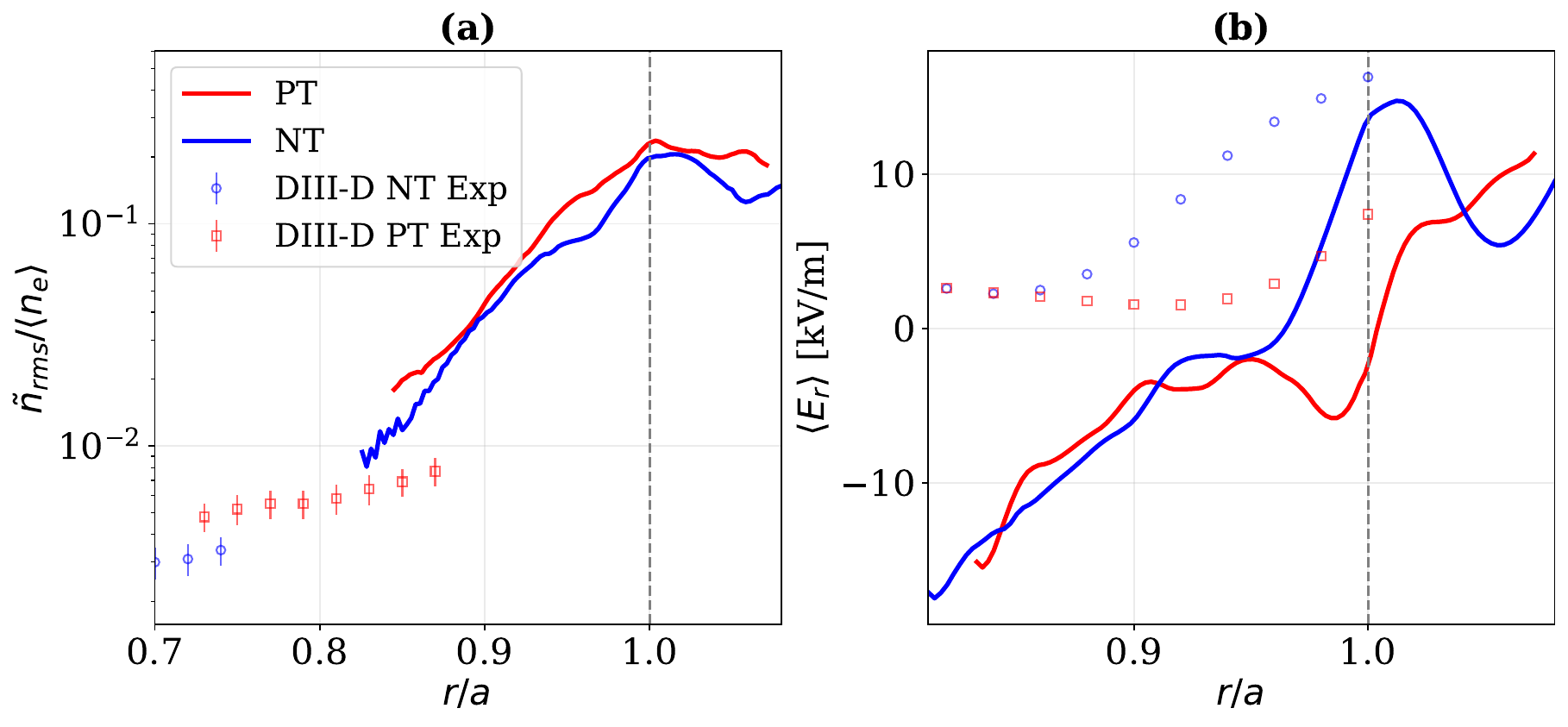}
    \caption{Time- and y-averaged radial profiles at the OMP of (a) relative fluctuation levels for electron density and (b) radial electric field comparing the PT (red) and NT (blue) configurations. Both include available DIII-D experimental measurements as scatter points. The vertical dashed line marks the LCFS.}
    \label{fig:nt-pt-dn-Er}
\end{figure}

Comparison with other experimental data captures trends relevant to turbulent transport. Figure~\ref{fig:nt-pt-dn-Er}(a) shows the RMS density-fluctuation amplitude normalized by the mean density. Simulated fluctuations are defined as the total quantity minus its $y$-averaged value ($\tilde{a} = a - \langle a \rangle_y$). The values shown are evaluated at the OMP and normalized by the time- and $y$-averaged baseline quantity. Beam emission spectroscopy (BES) measurements are available over a limited radial range. The simulations reproduce the experimental trend that density fluctuations are larger in PT than in NT.

Figure~\ref{fig:nt-pt-dn-Er}(b) compares the simulated OMP radial-electric-field profiles with values inferred from CER measurements. The simulations capture several experimental trends: the field magnitude at the LCFS is larger in NT than in PT, and the PT profile has a local minimum closer to the LCFS. Other discrepancies may result from the limited radial extent or the low-order Miller approximation, which becomes less accurate away from the LCFS. Nonetheless, the differences in radial electric field correlate with substantial differences in turbulent transport, as discussed in Sec.~\ref{subsec:turb-sup}.

\subsection{Turbulence suppression mechanisms} \label{subsec:turb-sup}
\begin{figure}
    \centering
    \includegraphics[width=0.99\linewidth]{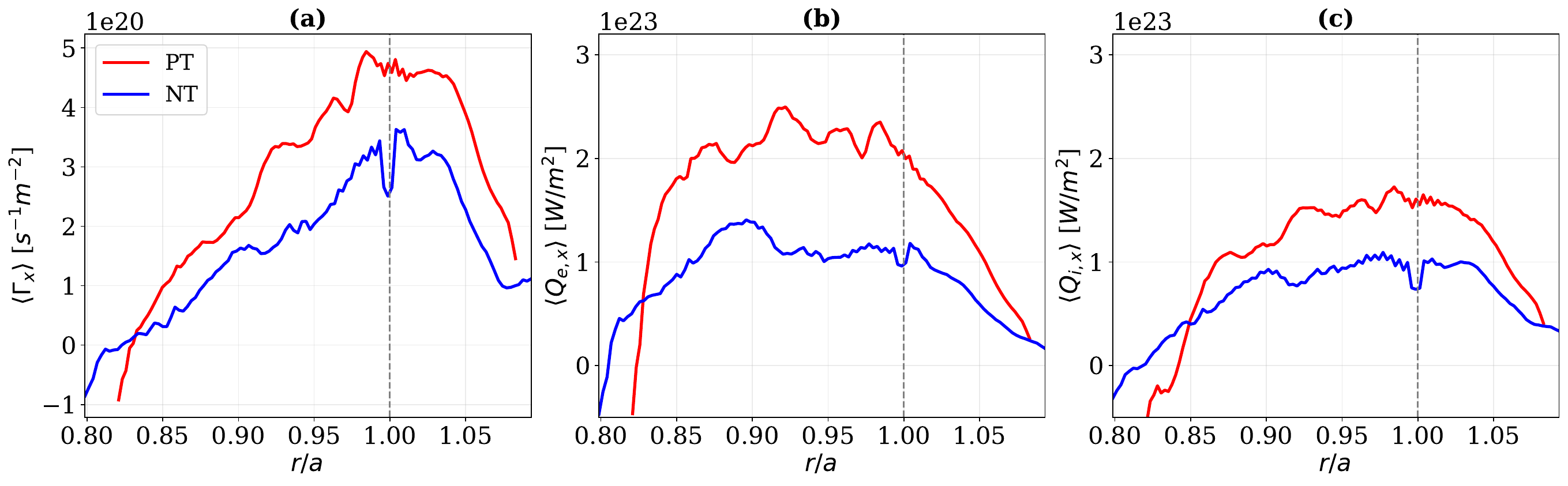}
    \caption{Time- and y-averaged radial profiles at the OMP of cross-field transport fluxes displaying (a) radial particle flux, (b) radial electron heat flux, and (c) radial ion heat flux for the PT (red) and NT (blue) configurations. The vertical dashed line marks the LCFS.}
    \label{fig:nt-pt-fluxes}
\end{figure}

We next characterize the turbulence-suppression mechanism in NT. Figure~\ref{fig:nt-pt-fluxes} shows that NT reduces the OMP radial particle flux and both electron and ion heat fluxes across most of the domain. In NT, the electron and ion heat fluxes are comparable; in PT, the electron heat flux is more than twice the ion heat flux.

\begin{figure}
    \centering
    \includegraphics[width=0.99\linewidth]{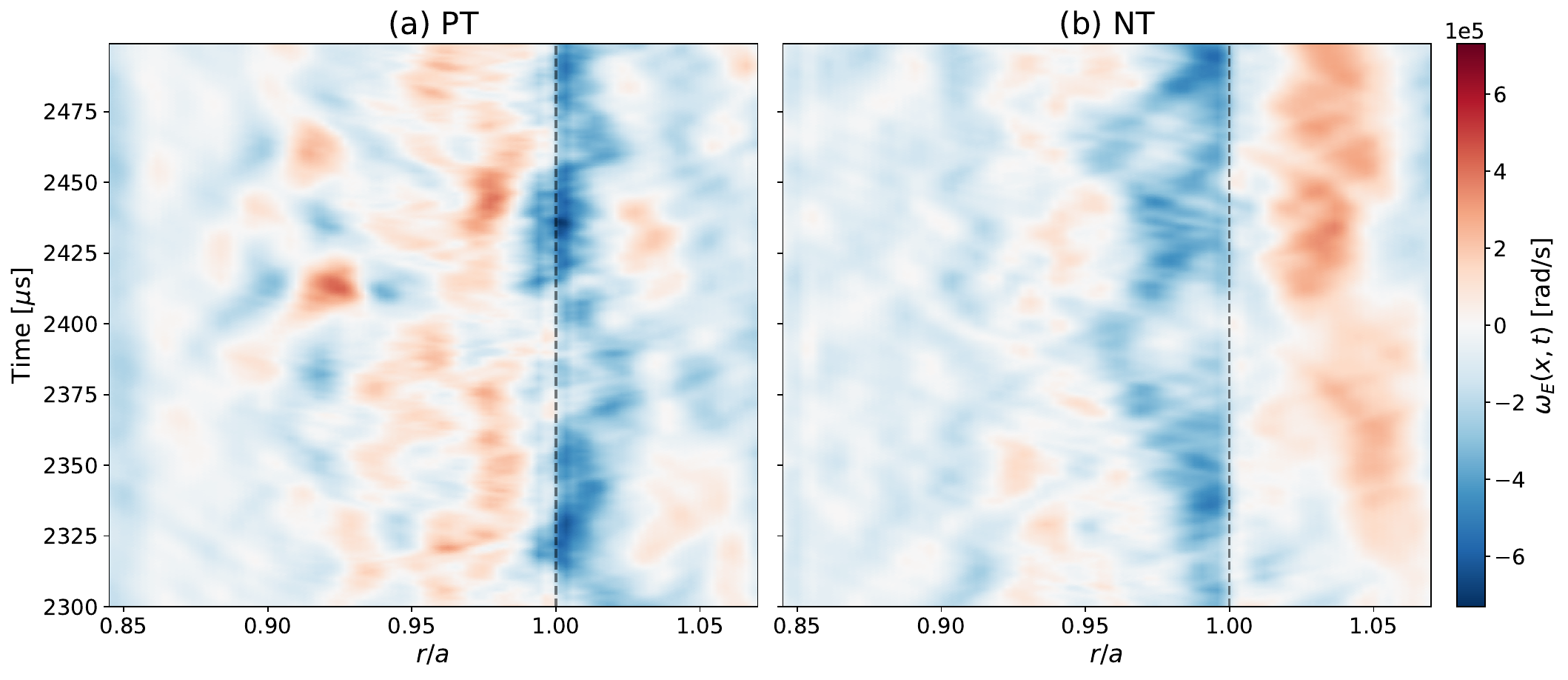}
    \caption{Spatiotemporal evolution of the zonal shearing rate $\omega_E(x, t)$ across the edge and scrape-off layer for the (a) PT and (b) NT configurations. The vertical dashed line denotes the position of the LCFS.}
    \label{fig:zf-time-space}
\end{figure}

Figure~\ref{fig:zf-time-space} shows the spatiotemporal evolution of the zonally averaged $E \times B$ radial shearing rate, $\omega_E(x, t) = \partial \langle V_{E,y} \rangle_y / \partial x$. The NT configuration in panel (b) exhibits a temporally coherent band inside the LCFS, indicating robust low-frequency zonal flows. In contrast, the PT shearing structures are transient and fragmented, changing sign in both radius and time.

\begin{figure}
    \centering
    \includegraphics[width=0.66\linewidth]{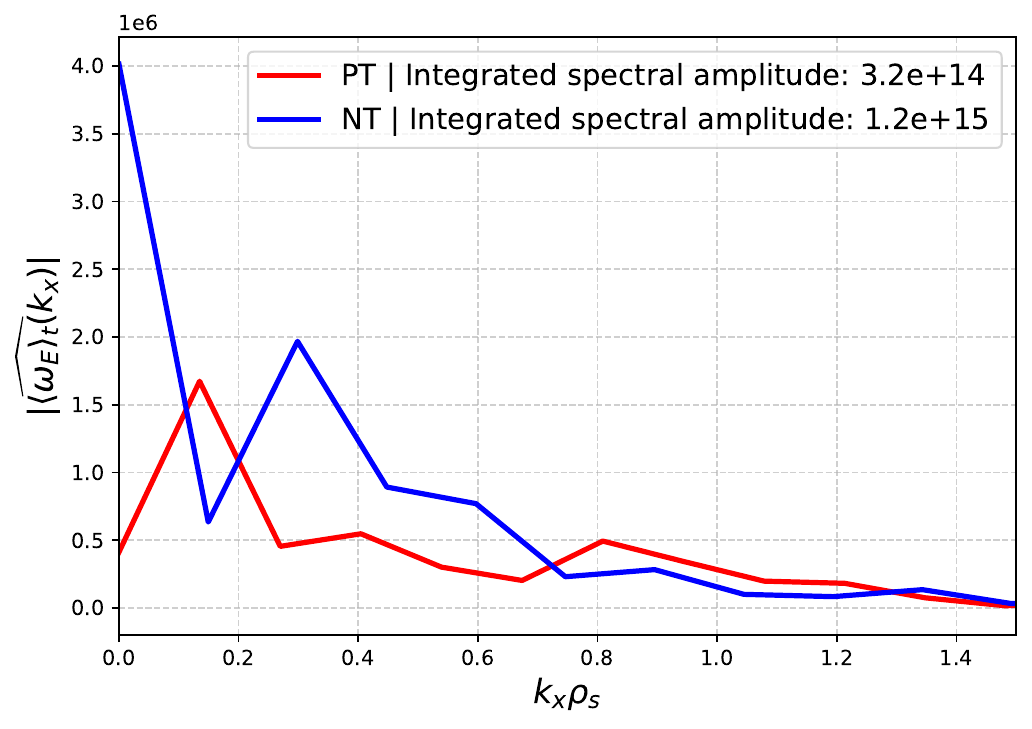}
    \caption{Radial-wavenumber ($k_x \rho_s$) spectrum of the time-averaged, zero-frequency zonal shear amplitude $|\langle \omega_E \rangle_t|$ over $0.85 \leq r/a \leq 1.0$ for the PT (red) and NT (blue) configurations, with the integrated spectral amplitude reported in the legend.}
    \label{fig:zf-zerofreq}
\end{figure}

Figure~\ref{fig:zf-zerofreq} displays the zero-frequency radial wavenumber spectrum of the zonal shearing rate, obtained by time averaging, applying a Tukey window, and performing a one-dimensional radial Fourier transform over $0.85 \leq r/a \leq 1.0$. Both configurations exhibit shear over a broad range of radial scales, but the NT spectrum is generally larger, particularly at the lowest radial wavenumbers, consistent with the triangularity comparisons in Ref.~\cite{singh2025}. The integrated spectral amplitude is approximately 3.75 times as large in NT as in PT ($1.2\times10^{15}$ versus $3.2\times10^{14}$), demonstrating stronger stationary zonal shear inside the NT LCFS.

\begin{figure}[htbp]
    \centering
    \includegraphics[width=0.8\textwidth]{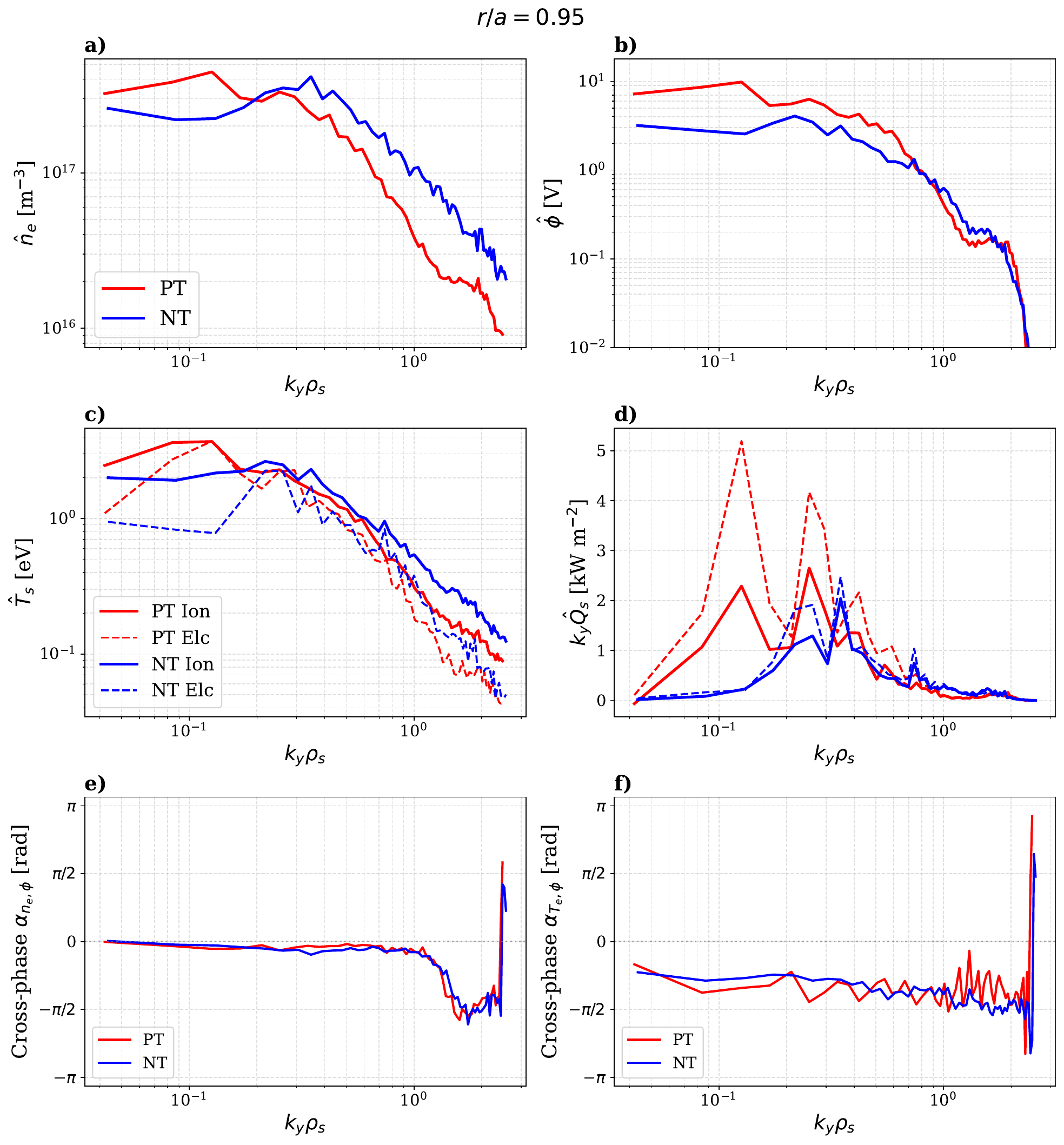}
    \caption{Binormal wavenumber spectra evaluated at $r/a = 0.95$ comparing PT (red) and NT (blue) configurations: (a) RMS electron density fluctuations $\hat{n}_e$, (b) electrostatic potential fluctuations $\hat{\phi}$, (c) ion (solid) and electron (dashed) temperature fluctuations $\hat{T}_s$, (d) the spectral density of the total turbulent cross-field heat flux $k_y \hat{Q}_s$, (e) density cross-phase, and (f) electron temperature cross-phase.}
    \label{fig:fsa_multi_field_spectra}
\end{figure}

To investigate how this enhanced shear modulates individual transport channels, we examine the flux-surface-averaged poloidal-wavenumber ($k_y \rho_s$) spectra of the turbulent fluctuations, resulting cross-field heat fluxes, and cross-phases in Figure~\ref{fig:fsa_multi_field_spectra}. These quantities are evaluated at $r/a = 0.95$, where the difference in fluxes is largest in Fig.~\ref{fig:nt-pt-fluxes}. At low $k_y\rho_s$, the electron-density fluctuation spectrum in panel (a) is lower in NT; at higher wavenumbers, it is larger. Potential fluctuations in panel (b) are lower in NT across the full spectrum. Temperature fluctuations in panel (c) follow a similar scale-dependent trend, being slightly lower in NT at low wavenumbers and higher at large wavenumbers. Despite these comparable, and in some cases larger, fluctuation amplitudes, the turbulent heat-flux spectra in panel (d) are substantially lower in NT, particularly at the low-$k_y$ scales that dominate transport.

The cross-phase spectra in Figure~\ref{fig:fsa_multi_field_spectra}(e)--(f) provide additional insight into the reduced transport efficiency in NT. Over much of the transport-dominant wavenumber range, the density-potential cross-phase, \(\alpha_{n_e\phi}\), in panel (e) remains close to zero in both configurations. Because the radial \(E\times B\) velocity is proportional to \(ik_y\tilde{\phi}\), the turbulent particle flux scales as \(k_y|\tilde{n}_e||\tilde{\phi}|\sin(\alpha_{n_e\phi})\). Fluctuations with a near-zero cross-phase therefore contribute little to radial transport despite their finite amplitudes, so the larger high-$k_y\rho_s$ density-fluctuation power in NT does not produce correspondingly greater transport. By contrast, the $T_e$--$\phi$ cross-phase in panel (f) is nonzero in both configurations and approaches $-\pi/2$. Together with the greater low-wavenumber temperature-fluctuation power in PT, this phase relationship produces the larger PT heat-flux spectrum shown in panel (d).

\begin{figure}[htbp]
    \centering
    \includegraphics[width=0.8\textwidth]{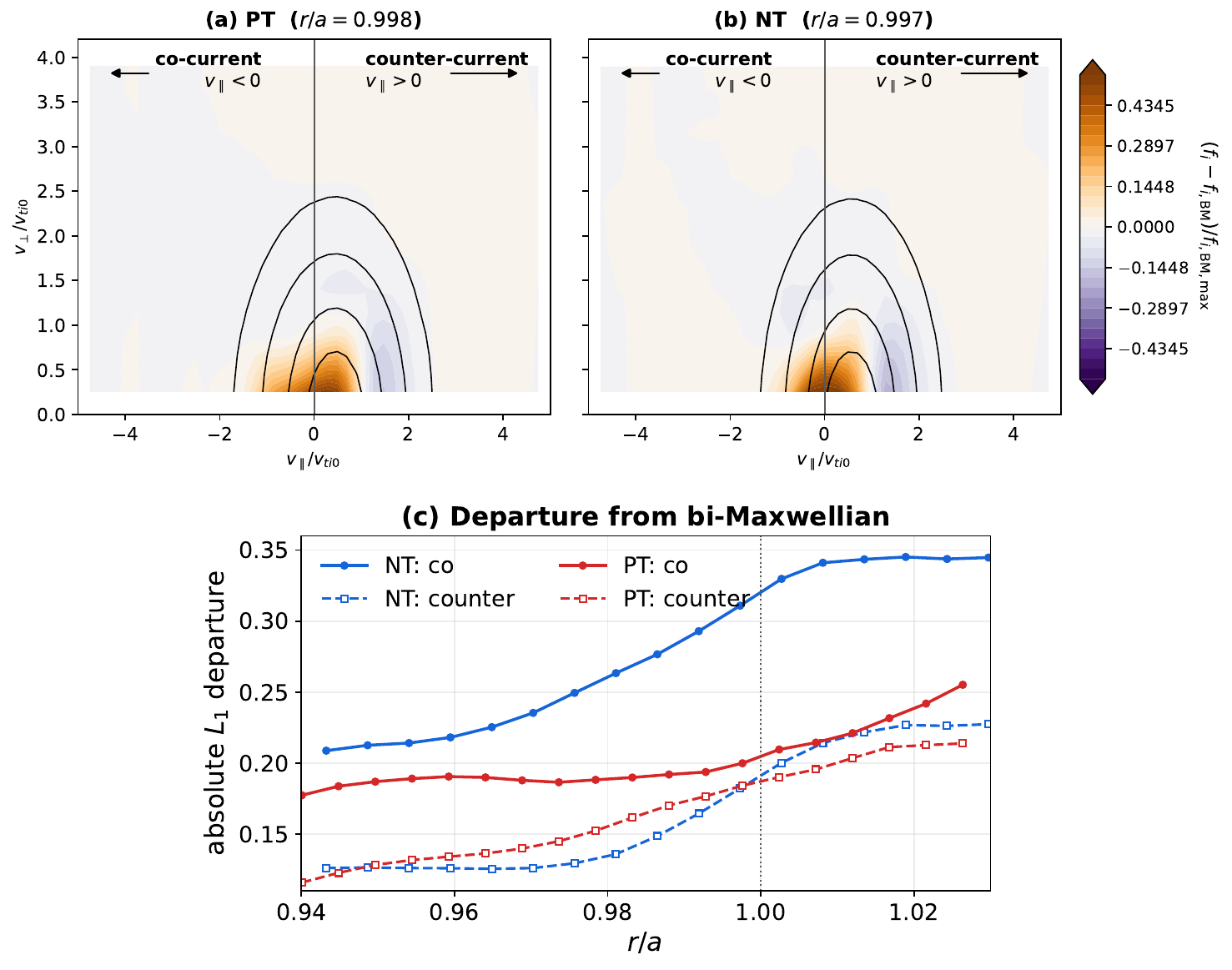}
    \caption{Ion kinetic signatures at $t=2.4~\mathrm{ms}$. Panels (a) and (b) show the difference of ion distribution functions from a locally matched bi-Maxwellian for PT and NT, respectively. Panel (c) is the directional $L_1$ departure from the bi-Maxwellian, as defined in Eq.~\eqref{eq:directional-L1}. Solid and dashed curves denote co- and counter-current ions, respectively; the dotted line marks the LCFS.}
    \label{fig:ion-kinetic-effects}
\end{figure}

The differences in zonal shear are accompanied by distinct ion kinetic behavior near the LCFS. Panels (a) and (b) of Figure~\ref{fig:ion-kinetic-effects} show, near the LCFS at 2.4~ms, the differences between the ion distribution functions and locally matched bi-Maxwellians. Both geometries exhibit clear distortions. We quantify these velocity-selective departures using the normalized directional measure
\begin{equation}
L_{1,d}(r)=
\frac{\displaystyle
\int_{\mathcal{D}_d}
\left|
\left\langle f_i \right\rangle_y-f_{i,\mathrm{BM}}
\right|
\,\mathrm{d}v_\parallel\,\mathrm{d}\mu}
{\displaystyle
\int_{\mathcal{D}_d}
f_{i,\mathrm{BM}}
\,\mathrm{d}v_\parallel\,\mathrm{d}\mu},
\label{eq:directional-L1}
\end{equation}
where $\left\langle f_i\right\rangle_y$ is evaluated at the outboard midplane and $f_{i,\mathrm{BM}}$ is constructed from the density, parallel flow, and parallel and perpendicular temperatures averaged over $y$ at the same location. For the convention
used here, $\mathcal{D}_{\mathrm{co}}=\{v_\parallel<0\}$ and $\mathcal{D}_{\mathrm{counter}}=\{v_\parallel>0\}$. Figure~\ref{fig:ion-kinetic-effects}(c) reveals a substantially larger
co-current departure in NT near the plasma edge. This asymmetry is qualitatively consistent with finite-orbit-width and ion-orbit-loss effects, which can generate strong edge radial electric fields \cite{StoltzfusDueck2012,StoltzfusDueck2020,Zhu2023_IOL} and may be particularly important in NT configurations \cite{Nishimura2020,Wang2023,kramer2025}. A detailed analysis of shear formation in these geometries, including the effect of velocity-selective orbit physics, will be presented in a future publication.

\section{Conclusions and Discussion}
\label{sec:conclusions}
In this work, we present global full-$f$ gyrokinetic simulations of DIII-D inner-wall-limited plasmas using Gkeyll to investigate edge and SOL transport in PT and NT configurations. A key advance is the incorporation of a reduced neutral-fluid model that directly includes ionization sources and charge-exchange momentum transfer in the gyrokinetic evolution. Relative to an adaptive recycling-source treatment, the neutral model improves particle and energy balances, produces smoother interior flow structures while preserving the dominant edge shearing layer, and yields better agreement with measured radial density and temperature profiles.

The comparison between PT and NT reveals distinct edge-transport dynamics. Although fluctuation amplitudes remain broadly comparable, NT exhibits substantially reduced turbulent transport, stronger and more persistent stationary zonal shear, and near-zero Reynolds stress in the region of strongest shear. These results indicate that turbulence is not the primary mechanism sustaining the observed radial electric field. Kinetic analysis reveals a larger co-current directional $L_1$ departure (Eq.~\ref{eq:directional-L1}) from a locally matched bi-Maxwellian in NT near the plasma edge. Together, these signatures indicate an underlying velocity-selective kinetic process that may contribute to the edge radial electric field and stationary zonal shear. Finite-orbit-width and ion-orbit-loss effects are plausible explanations; future work will determine which mechanisms dominate shear formation.

Additional ongoing work will focus on improving the quantitative treatment of the experimental power balance and simulation initialization. The remaining discrepancy in the PT electron-temperature profile may arise because the simulations do not reproduce the experimentally inferred edge power deposition. Incorporating realistic ECH deposition will require axisymmetric simulations evolved over tens of milliseconds to establish the background plasma, followed by three-dimensional restarts to resolve turbulent transport. The reduced neutral-fluid model will also be extended to evolve the neutral fluid dynamically, enabling a more complete treatment of neutral transport and plasma–neutral coupling across the edge and SOL.

Beyond the comparison of triangularities, this work demonstrates the value of treating the confined plasma, SOL, sheath boundary conditions, and neutral interactions within a unified global gyrokinetic framework. This framework provides a foundation for predictive studies of confinement and boundary transport in ELM-free regimes and for extrapolating these regimes to reactor-relevant conditions.

\section*{Acknowledgements}
T. Bernard acknowledges helpful discussions with C.F.B. Zimmermann, the DIII-D Negative Triangularity Team, and CEDA SciDAC collaborators, and thanks the Gkeyll Team for code development support.

This research used resources of the National Energy Research Scientific Computing Center (NERSC), a Department of Energy User Facility using NERSC awards FES-ERCAP0036858, FES-ERCAP0037492, and FES-ERCAP0036584.

This material is based upon work supported by the U.S. Department of Energy (DOE), Office of Science, Office of Fusion Energy Sciences, through the CEDA SciDAC project under Award No.~DE-AC02-09CH11466 (Princeton Plasma Physics Laboratory Field Work Proposal No.~3219). This material is also based upon work supported by the U.S. Department of Energy, Office of Science, Office of Fusion Energy Sciences, using the DIII-D National Fusion Facility, a DOE Office of Science user facility, under Awards DE-FC02-04ER54698 and DE-FG02-08ER54999. A.C.D. Hoffmann, M. Francisquez, and G.W. Hammett were additionally supported by a Distinguished Scientist award and other PPPL projects under DOE Contract No.~DE-AC02-09CH11466.

This report was prepared as an account of work sponsored by an agency of the United States Government. Neither the United States Government nor any agency thereof, nor any of their employees, makes any warranty, express or implied, or assumes any legal liability or responsibility for the accuracy, completeness, or usefulness of any information, apparatus, product, or process disclosed, or represents that its use would not infringe privately owned rights. Reference herein to any specific commercial product, process, or service by trade name, trademark, manufacturer, or otherwise, does not necessarily constitute or imply its endorsement, recommendation, or favoring by the United States Government or any agency thereof. The views and opinions of authors expressed herein do not necessarily state or reflect those of the United States Government or any agency thereof.

\appendix

\section{Detailed model equations} \label{app:model-eqns}
Additional details of the collision operator, adaptive source model, and neutral dynamics are contained in this appendix.

\subsection{Dougherty collisions} \label{app:coll}
We model plasma collisions using a Dougherty operator generalized to multi-species interactions~\cite{Francisquez2020lbo,francisquez2022improved}. This operator takes the form
\begin{equation}
    \jac_s C[f_s] = \sum_r \nu_{sr} \nabla_v \cdot [(\v{v} - \v{u}_{rs})\jac_sf_s + v_{t,sr}^2 \nabla_v \jac_s f_s ],
\end{equation}
where $\nu_{sr}$ is the collision frequency for species $s$ with species $r$. The cross-species primitive moments $\v{u}_{rs}$ and $v_{t,sr}^2$ are constructed to ensure exact conservation of momentum and energy \cite{francisquez2022improved}. Consistent with the global nature of these simulations, the collision frequencies vary spatiotemporally according to the calculated species moments:
\begin{equation}
    \nu_{sr}(\v{R},t) = \nu_{sr0} \frac{n_r(\v{R},t)}{n_{r0}}\left[ \frac{v_{ts0}^2 + v_{tr0}^2}{v_{ts}^2(\v{R},t) + v_{tr}^2(\v{R},t)}\right]^{3/2}, \label{eq:cross-coll}
\end{equation}
with the reference collision frequency given by
\begin{equation}
    \nu_{sr0} = \frac{1}{m_s}\left(\frac{1}{m_s} + \frac{1}{m_r} \right)\frac{q_s^2 q_r^2 \log \Lambda_{sr}}{3(2\pi)^{3/2}\epsilon_0^2}\frac{n_{r0}}{(v_{ts0}^2 + v_{tr0}^2)^{3/2}}, \label{eq:ref-coll}
\end{equation}
where $n_{r0}$, $v_{tr0}^2$, and $v_{ts0}^2$ are prescribed background quantities, and $\log \Lambda_{sr}$ is the Coulomb logarithm.

\subsection{Neutral collisions} \label{app:neutrals}
Here we describe the atomic and molecular collision operators incorporated into Gkeyll. They enter the right-hand side of the gyrokinetic equation (\ref{eq:gk-eqn}) through $C[f_s]$. Because these operators readily generalize to multi-species impurity states, the notation applies to both neutral and plasma species. We extend the ionization model presented in Ref.~\cite{bernard2022} to distinguish the scattered primary-electron population from the newly ejected secondary-electron population. This distinction accounts for the fact that ejected electrons are typically born cold relative to the scattered primary electrons. The electron ionization collision operator is
\begin{equation}
    C_{iz}^e = n_{d} \langle v_e \sigma \rangle_{iz} [f_{e1,M}(n_e, u_{\parallel,e}, v^2_{t,iz1}) +  f_{e2,M}(n_e, u_{\parallel,d}, v^2_{t,iz2}) - f_e],
\end{equation}
where subscripts 1 and 2 denote the scattered and ejected electron populations, respectively; the subscript $d$ denotes the donor species (neutral or ion) impacted by the incident electron; and the index $M$ indicates that Maxwellian distributions model the resulting populations. The associated thermal speeds are defined via $v^2_{t,iz1} = T_{e1}/m_e$ and $v^2_{t,iz2} = T_{e2}/m_e$, where the ejected electron temperature is determined from
\begin{equation}
    T_{e2} =
    \begin{cases}
        \frac{1}{3} \sqrt{E_{iz} \left( \frac{3}{2} T_e - E_{iz} \right)} & \text{for } \frac{3}{2} T_e - E_{iz} \geq E_{iz} \\
        \frac{1}{3} \left( \frac{3}{2} T_e - E_{iz} \right) & \text{for } \frac{3}{2} T_e - E_{iz} < E_{iz}
    \end{cases}
\end{equation}
and $E_{iz}$ is the ionization energy threshold parameterized in Ref.~\cite{voronov1997}. To avoid unphysical cooling, we enforce a temperature floor via $T_{e2} = \max(T_{e2}, T_{\rm floor})$, where $T_{\rm floor} = 3$~eV. The effective temperature of the scattered primary electrons is then evaluated as
\begin{equation}
    T_{e1} = \max(T_e - 2E_{iz}/3 - T_{e2}, T_{\rm floor} ).
\end{equation}
Total energy conservation is strictly maintained provided both $T_{e1}$ and $T_{e2}$ remain above $T_{\rm floor}$. The corresponding ionization collision terms for the newly born ion and the depleted donor species are:
\begin{eqnarray}
    C_{iz}^i &= n_e \langle v_e \sigma \rangle_{iz} f_{M,d} \\
    C_{iz}^d &= -n_e \langle v_e \sigma \rangle_{iz} f_d,
\end{eqnarray}
where $\langle v_e \sigma \rangle_{iz}$ represents the OpenADAS ionization rate coefficient. The recombination collision terms are written as:
\begin{eqnarray}
    C^{e}_{rec} &= -n_i \langle v \sigma \rangle_{rec} f_e \\
    C^{i}_{rec} &= -n_e \langle v \sigma \rangle_{rec} f_i \label{eq:rec-i}\\
    C^{r}_{rec} &= n_e \langle v \sigma \rangle_{rec} f_{M,i}, \label{eq:rec-r}
\end{eqnarray}
where the superscript $r$ in Eq.~(\ref{eq:rec-r}) indicates the receiver species capturing the electron (either a fluid neutral or a kinetic ion state). The receiver distribution $f_{M,i}$ is a Maxwellian defined by the local ion moments, accommodating scenarios where this collision term is projected onto a phase-space grid distinct from that of the source ion in Eq.~(\ref{eq:rec-i}). The recombination rate coefficient $\langle v \sigma \rangle_{rec}$ is likewise extracted from the OpenADAS database.

The full resonant charge exchange interaction between ions and neutrals of equal mass is described by the Boltzmann integral:
\begin{eqnarray}
    C^{cx}_i = -C^{cx}_n = &f_n(\mathbf{v}) \int f_i(\mathbf{v}') \sigma_{cx}(|\mathbf{v} - \mathbf{v}'|) |\mathbf{v} - \mathbf{v}'| \, d\mathbf{v}'  \nonumber \\
    &- f_i(\mathbf{v}) \int f_n(\mathbf{v}') \sigma_{cx}(|\mathbf{v} - \mathbf{v}'|) |\mathbf{v} - \mathbf{v}'| \, d\mathbf{v}', \label{eq:cx-full}
\end{eqnarray}
To avoid evaluating this computationally intensive velocity-space integral at every time step, we implement a simplified algebraic model based on Ref.~\cite{meier2011}. The reduced charge exchange operators become
\begin{equation}
    C^{cx}_i = -C^{cx}_n = \sigma_{cx} V_{cx} (n_i f_n - n_n f_i), \label{cx-in}
\end{equation}
where $V_{cx}$ is the effective relative velocity defined as
\begin{equation}
    V_{cx}^2 \equiv \frac{4}{\pi}(v_{t,i}^2 + v_{t,n}^2) + (\mathbf{u}_i - \mathbf{u}_n)^2.
\end{equation}
We evaluate the energy-dependent charge exchange cross sections using standard analytical fits from Ref.~\cite{Barnett1990}, considering reactions for hydrogen ($\text{H}\text{-}\text{H}^+$), deuterium ($\text{D}\text{-}\text{D}^+$), helium, and neon via the logarithmic relation
\begin{eqnarray}
    \sigma_{cx, X} = a_X - b_X \ln(V_{cx}).
\end{eqnarray}

\subsection{Adaptive source} \label{app:adpt-src}
We use the adaptive source model presented in \cite{Hoffmann2026_AdptSrc}. The phase-space volumetric source term $S_s$ in the gyrokinetic equation (\ref{eq:gk-eqn}) is the sum of $k$ sub-sources $S_s^k$. Each sub-source is defined by a non-drifting Maxwellian in velocity space and a Gaussian distribution in configuration space:
\begin{equation}
    S_s^k = G^k_s(\v{R}; \Gamma_s^k) M^k_s[\vpar, \mu; T_s^k],
\end{equation}
where the time-dependent scalars $\Gamma_s^k$ and $T_s^k$ represent the particle source rate and source temperature, respectively. The spatial profile is defined as
\begin{equation}
    G^k_s(\v{R}; \Gamma_s^k) = \Gamma_s^k \eta_s^k(\v{R}),
\end{equation}
with the normalized profile given by
\begin{equation}
\eta_s^k(\v{R}) = C_s^k \prod_{i=x,z}\exp{\left[ -\frac{(x_i - \xi_{i,s,k})^2}{2\sigma_{i,s,k}^2}\right]}. \label{eq:unit-Gaussian}
\end{equation}
The scalar $C_s^k$ enforces the normalization condition $\int \eta_s^k d\v{R} = 1$. The non-drifting Maxwellian takes the form
\begin{equation}
    M^k_s[\vpar, \mu; T_s^k] = \left( \frac{m_s}{2 \pi T_s^k} \right) \exp \left[ - \frac{m_s \vpar^2}{2 T_s^k} - \frac{\mu B}{T_s^k}\right].
\end{equation}

We dynamically adjust $\Gamma_s^k$ and $T_s^k$ to maintain the user-defined input particle rate $\Gamma_{k,s}^{in}$ and input power $P_{k,s}^{in}$. These quantities compensate for the instantaneous particle ($\dot{N}_r^k$) and energy ($\dot{E}_r^k$) losses of species $r$ across the boundaries. At each time step, they are updated according to
\begin{eqnarray}
    \Gamma_s^k(t_{n+1}) &= \Gamma_{k,s}^{in} + \dot{N}_r^k(t_n) \\
    T_s^k(t_{n+1}) &= \frac{2}{3}\frac{P^{in}_{k,s} + \dot{E}_r^k(t_n)}{\Gamma_s^k(t_{n+1})}.
\end{eqnarray}
A sub-source for species $s$ can compensate for the losses of a distinct species $r$ to model ambipolar processes, such as recycling, when omitting explicit neutral collisions.

\section*{References}
\bibliographystyle{iopart-num}
\bibliography{references}

\end{document}